\documentclass[twocolumn,twoside,slac_two]{revtex4}
\usepackage{graphicx}
\usepackage{fancyhdr}
\bibpunct{[}{]}{;}{a}{}{,}
\newcommand\ATel{ATel \# }

     \newcommand{\la}{\,\rlap{\raise 0.5ex\hbox{$<$}}{\lower 1.0ex\hbox{$\sim$}}\,}
     \newcommand{\ga}{\,\rlap{\raise 0.5ex\hbox{$>$}}{\lower 1.0ex\hbox{$\sim$}}\,}

\def\aap {Astron. \& Astrophys.\/}

\def\apj {Astrophys. J.\/}
\def\apjl {Astrophys. J. Letters}

\def\mnras {Mon. Not. R. Astron. Soc.\/}

\begin{document}

\title{Properties of Supergiant Fast X-ray Transients as observed by \emph{Swift} }


\author{P.\ Romano, S.\ Vercellone}
\affiliation{INAF-IASF Palermo, Via Ugo La Malfa 153, Palermo, Italy}
\author{H.A.\ Krimm}
\affiliation{NASA/Goddard Space Flight Center, Greenbelt, MD 20771, USA}

\author{P.\ Esposito}
\affiliation{INAF-OA Cagliari,  localit\`a Poggio dei Pini, 
     Strada 54, I-09012 Capoterra, Italy}

\author{G.\ Cusumano, V.\ La Parola, V.\ Mangano}
\affiliation{INAF-IASF Palermo, Via Ugo La Malfa 153, Palermo, Italy   }

\author{J.A.\ Kennea, D.N.\ Burrows}
\affiliation{Department of Astronomy and Astrophysics, Pennsylvania State 
        University, University Park, PA 16802, USA}

\author{C.\ Pagani}
\affiliation{Department of Physics \& Astronomy, University of Leicester, LE1 7RH, UK}

\author{N.\ Gehrels}
\affiliation{NASA/Goddard Space Flight Center, Greenbelt, MD 20771, USA }

\begin{abstract}
We present the most recent results from our investigation on 
Supergiant Fast X-ray Transients, a class of High-Mass X-ray Binaries, 
with a possible counterpart in the gamma-ray energy band. 
Since 2007 {\it Swift} has contributed to this new field 
by detecting outbursts from these fast transients with the BAT and by
following them for days with the XRT. Thus, we demonstrated that while 
the brightest phase of the outburst only lasts a few hours, further 
activity is observed at lower fluxes for a remarkably longer time, up to 
weeks. Furthermore, we have performed several campaigns of intense monitoring 
with the XRT, assessing the fraction of the time these sources spend in each 
phase, and their duty cycle of inactivity. 
\end{abstract}

\maketitle

\thispagestyle{fancy}


\begin{figure*}[t]%%%%%%%%%%%%%%%%%%%%%%%%%%%%%%%%%%%%%%%%%%%%%
\centering
\includegraphics[height=17cm,width=5cm,angle=270]{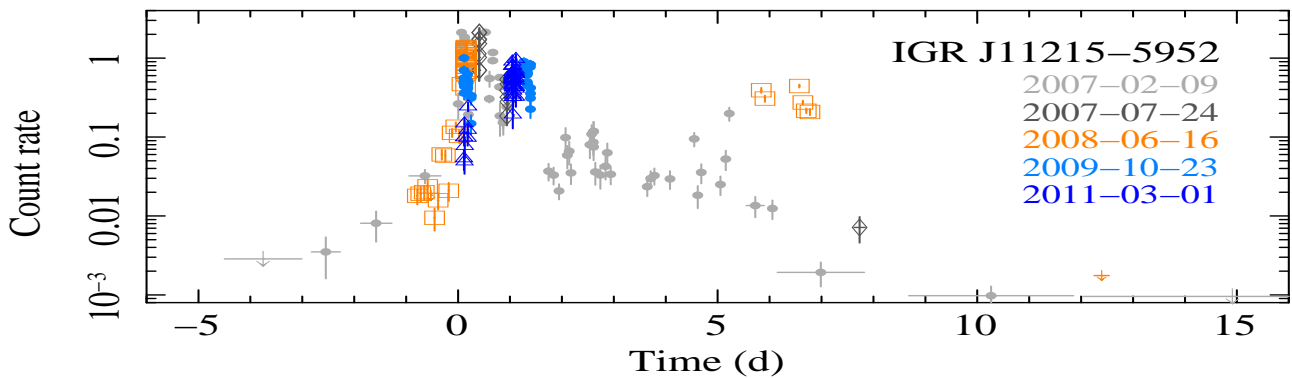}
\caption{{\it Swift}/XRT light curves of the  2007 Feb 9 outbursts (light grey filled circles), 
	        2007 July 24 (dark grey empty diamonds), 2008 June 16 (orange empty squares), 
                2009 November 23 (cyan filled circles), and 2011 March 1 (blue empty triangle)
                folded with a period of 164.6 days,
                relative to the peak of the first outburst.  
Data from \cite{Romano2007,Sidoli2007,Romano2009:11215_2008,Esposito2009:atel2257,Romano2011:atel3200}.} 
\label{fermi11:fig:11215}
\end{figure*}%%%%%%%%%%%%%%%%%%%%%%%%%%%%%%%%%%%%%%%%%%%%%

\begin{table*}[t]%%%%%%%%%%%%%%%%%%%%%%%%%%%%%%%%%%%%%%%%%%%%%  Table 1 
\begin{center}
\begin{tabular}{lccccccc}
\hline
\hline
Name             &Campaign Dates                &Obs.\     &Expo. &$\Delta T_{\Sigma}$  & $P_{\rm short}$ &  IDC  & Rate$_{\Delta T_{\Sigma}}$   \\
%    		                                       &  &N. & (ks) &(ks) & (\%) &  (\%) &  	\\ 	
    		                                             & & & (ks) &(ks) & (\%) &  (\%) &   ($10^{-3}$counts s$^{-1}$)	\\ 	
\hline
IGR~J16479--4514 &  2007-10-26 2009-11-01	& 144&	161 & 29.7 &3  & 19 & $3.1\pm0.5$ \\	      
XTE~J1739--302 	 &  2007-10-27 2009-11-01	& 184&	206 & 71.5 &10 & 39 & $4.0\pm0.3$ \\
IGR~J17544--2619 &  2007-10-28 2009-11-03	& 142&	143 & 69.3 &10 & 55 & $2.2\pm0.2$ \\
AX~J1841.0--0536 &  2007-10-26 2008-11-15	&  88&	 96 & 26.6 &3  & 28 & $2.4\pm0.4$ \\  
%\hline
 Total           &                             & 558&606 & & & & \\
\hline
\end{tabular}
\caption{The {\it Swift} long-term monitoring campaign (Figure~\ref{fermi11:fig:alllcvs}a--d). 
   $\Delta T_{\Sigma}$ is sum of the exposures accumulated in all observations (exposure $>900$\,s)  
   where only a 3-$\sigma$ upper limit was achieved;  
   $P_{\rm short}$ is the percentage of time lost to short observations; 
   IDC is the duty cycle of inactivity, i.e.,  
   the time each source spends undetected down to a flux limit of (1--3)$\times10^{-12}$ erg cm$^{-2}$ s$^{-1}$;
   Rate$_{\Delta T_{\Sigma}}$ is the cumulative count rate  (0.2--10\,keV). 
   Adapted from~\cite{Romano2009:sfxts_paperV,Romano2011:sfxts_paperVI}.  
\label{fermi11:tab:campaign} }
\end{center}
\end{table*}%%%%%%%%%%%%%%%%%%%%%%%%%%%%%%%%%%%%%%%%%%%%%

\section{INTRODUCTION \label{fermi11:intro}}

High mass X-ray binaries (HMXBs) are stellar systems composed of a 
compact object (CO) and an early-type massive star which are traditionally divided 
in two subclasses, depending on the nature of the high mass primary and 
the different mass-transfer and accretion mechanisms involved.  
The Be-HMXBs, are transient systems with main sequence Be primaries, 
in generally wide ($P_{\rm orb}\ga 10$\,d) eccentric ($e\sim 0.3$--0.5) orbits; 
in such systems the primaries are not filling their Roche lobes, 
and accretion onto the compact object occurs from the equatorial region 
of the rapidly rotating Be star.  
The X-ray emission from such systems is highly variable and mostly transient
(with a dynamic range of several orders of magnitude),
often with recurrent outbursts caused by an enhanced accretion rate 
when the compact star is close to periastron. 
OB supergiant HMXBs (sg-HMXBs), on the other hand, are systems with an 
evolved OB supergiant primary with with smaller ($P_{\rm orb}\la 10$\,d), 
more circular orbits than in Be-HMXBs. 
They are powered either by a geometrically thin accretion disc 
or by the strong radiation-driven stellar winds, depending on whether 
the primary fills its Roche lobe or not, and their X-ray emission 
is bright and persistent. 

Recently, a third class of HMXBs was added, the supergiant fast X-ray transients (SFXTs), 
which share characteristics with both of the above classes. 
SFXTs are associated with an O or B supergiant, but are are not persistent
sources, as they display outbursts characterized by bright flares lasting 
a few hours (as seen by INTEGRAL) with peak luminosities of 
10$^{36}$--10$^{37}$~erg~s$^{-1}$ 
\citep{Sguera2005,Sguera2006,Negueruela2006:ESASP604}, 
and a dynamic range of 3--5 orders of magnitude. 
These flares are significantly shorter than those of typical Be-HMXBs.
Their hard X-ray spectra resemble those of HMXBs 
hosting X-ray pulsars, a hard power law below 10\,keV, with a high 
energy cut-off at $\sim15$--30~keV, sometimes strongly absorbed at soft energies 
\citep{Walter2006}.   
Hence, it is generally assumed that all members of this class are 
HMXBs hosting a neutron star (NS), although pulse periods are measured only for 
half the sample, currently consisting of 10 confirmed members. 
About 20 more candidates are known which showed short transient flaring, 
but which have no confirmed association with an OB supergiant companion. 
Given the current interest in this class of sources, this number is bound to 
increase rapidly.
The actual causes of the bright outbursts are still being investigated,  
and they are probably related to either the properties of 
the wind from the supergiant companion 
\citep{zand2005,Walter2007,Negueruela2008,Sidoli2007} or to the 
presence of a centrifugal or magnetic barrier \citep{Grebenev2007,Bozzo2008}. 
The appeal of SFXTs originates from the properties they share with 
both classes in which HMXBs are traditionally divided, as they may represent  
an evolutionary connection between them. 
Recently, \cite{Sguera2008:11215,Sguera2009} proposed that the hard X-ray
counterparts of a few MeV unidentified EGRET, AGILE and {\it Fermi} transient sources
lasting only a few days, 3EG J1837$-$0423/HESS J1841$-$055, and 
EGR~J1122$-$5946, AGL~J2022$+$3622 may be associated with the SFXTs (or candidate)
sources  AX~J1841.0$-$0536, IGR~J11215$-$5952, and IGR~J20188$+$3647, respectively. 

\vfill

\section{IGR J11215$-$5952 \label{fermi11:11215}}

The hard X--ray transient source IGR J11215$-$5952 was discovered in April 2005 with 
INTEGRAL and is a confirmed SFXT. In 2007, archival INTEGRAL and RXTE observations had 
shown that the outbursts occurred with a periodicity of $\sim$330 days \citep{SidoliPM2006}, thus making 
IGR J11215$-$5952 the first SFXT displaying periodic outbursts, probably related to the orbital period. 
Taking advantage of this unprecedented property, we performed a target of opportunity (TOO) 
observation with {\it Swift}/XRT \citep{Burrows2005:XRT} to monitor the source around
the time of the next predicted outburst, on 2007 Feb 9.
We observed the source twice a day (2ks/day) from 2007 Feb 4 until Feb 9,
and then for $\sim 5$\,ks a day afterwards,
during a monitoring campaign that lasted 23 days for a total on-source 
exposure of $\sim73$\,ks. 
Figure~\ref{fermi11:fig:11215} shows how we could follow the source
for three orders of magnitude in flux, from non-detection up to the peak of the outburst at 
$10^{36}$ erg\,s$^{-1}$, and back down until it went below our detection limits. 
It also shows that the X-ray light curve is composed of several bright flares so that, 
while the bright outburst does last only a few hours, 
further significant activity (hence accretion onto the compact object) 
is seen at lower fluxes for a considerably longer (weeks) time.  
The brightest part of the outburst
lasted less than a day, on Feb 9, and would have been 
the only flaring activity seen with less sensitive instruments.
Furthermore, {\it Swift} allowed the determination of the orbital period of $\sim 164.6$\,d
 \cite{Sidoli2007,Romano2009:11215_2008},
a rare instance, as orbital periods are generally found from all-sky monitor data.

\begin{figure*}[t]%%%%%%%%%%%%%%%%%%%%%%%%%%%%%%%%%%%%%%%%%%%%%
\vspace{-3.5truecm}
\centering
\includegraphics[height=18cm,width=17cm,angle=0]{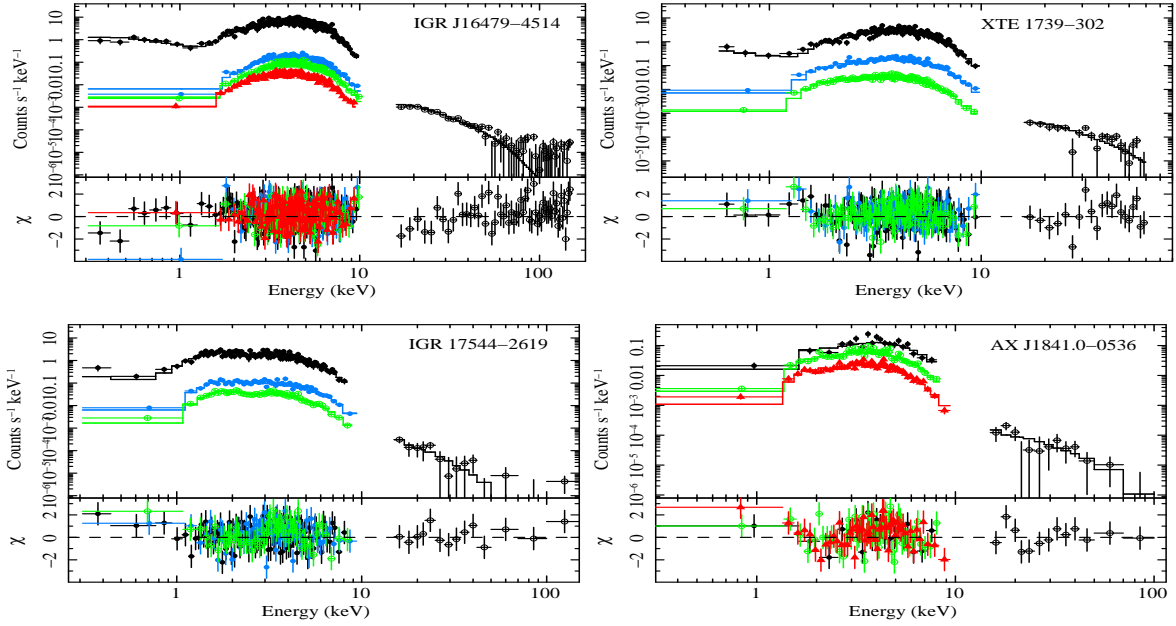}
\vspace{-6truecm}
\caption{Spectra of one representative outburst (black) of each of the four SFXTs monitored by {\it Swift}, 
fit with an absorbed power-law model with a high energy cut-off:   
filled circles and empty circles denote XRT and BAT data, respectively (see Table~\ref{fermi11:tab:spectra}, Sec.~\ref{fermi11:spectroscopy}). 
Data from \cite{Romano2008:sfxts_paperII,Sidoli2009:sfxts_paperIII,Romano2011:sfxts_paperVI,Romano2011:sfxts_paperVII}.
Intensity-selected spectra of out-of-outburst emission, fit with simple absorbed power-law models: 
Filled blue circles, green empty circles, and red filled triangles
            mark high, medium, and low states, respectively (see Table~\ref{fermi11:tab:spectra}, Sec.~\ref{fermi11:spectroscopy}). 
Data from \cite{Romano2009:sfxts_paperV,Romano2011:sfxts_paperVI}. 
} 
\label{fermi11:fig:outburst_spectra}
\end{figure*}%%%%%%%%%%%%%%%%%%%%%%%%%%%%%%%%%%%%%%%%%%%%%

\section{SWIFT'S SYSTEMATIC INVESTIGATION OF SFXTS \label{fermi11:swift}}

Since 2007, we have dedicated considerable {\it Swift} time to throughly and systematically
investigate the properties of SFXTs, with a strategy that combines 
monitoring programs with outburst follow-up observations. 
The most outstanding manifestation of the SFXT activity is indeed their outbursts, 
which {\it Swift} can catch and study broad band (0.3--150\,keV) in their early stages, 
thanks to its fast slewing and panchromatic capabilities. 
Furthermore, thanks to the flexible scheduling and low overheads, 
that make monitoring efforts cost-effective, we could give the first 
non-serendipitous attention to these objects with a high sensitivity X-ray telescope,
thus assessing spectroscopic and timing properties of SFXTs. 

The four targets we chose to monitor, IGR~J16479$-$4514, XTE~J1739--302, 
IGR~J17544$-$2619, and AX~J1841.0$-$0536, are all confirmed SFXTs, 
and include a source that had already triggered the BAT once, 
the two prototypes of the SFXT class, and a pulsar, respectively.  
Starting from 2007 October 26, we obtained 2--3 observations week$^{-1}$ object$^{-1}$, 
each 1\,ks long (see Table~\ref{fermi11:tab:campaign}), with XRT in AUTO mode, 
to best exploit XRT automatic mode switching \citep{Hill04:xrtmodes_mn} 
in response to changes in the observed fluxes.

\begin{figure*}[t]%%%%%%%%%%%%%%%%%%%%%%%%%%%%%%%%%%%%%%%%
\centering
\includegraphics[height=16.5cm,angle=90]{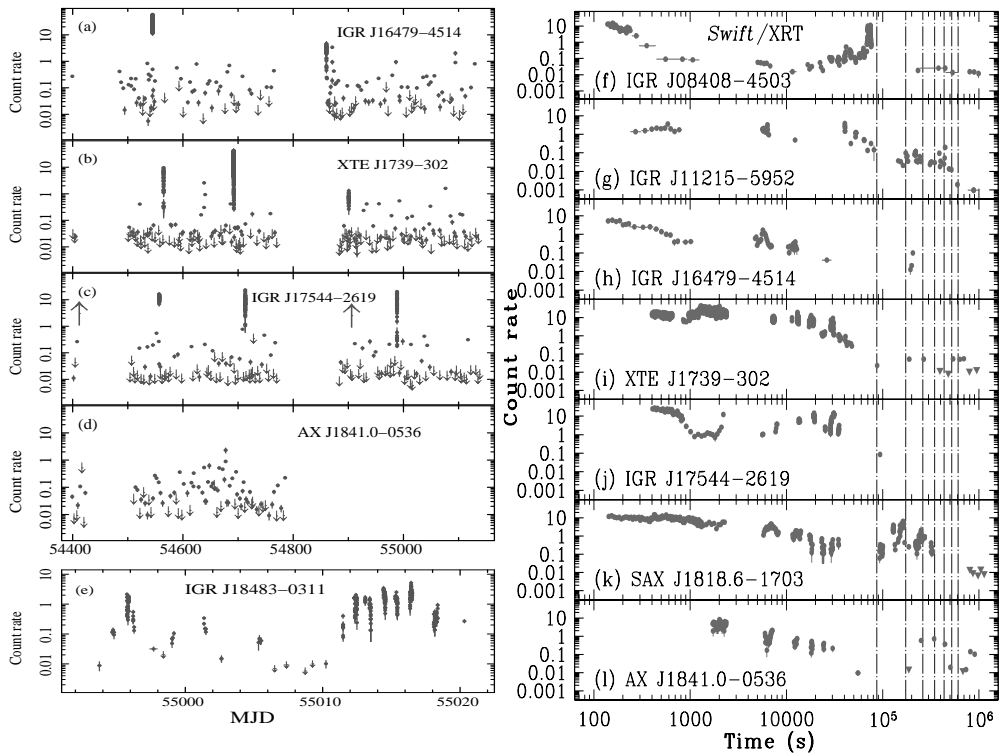}
\vspace{-1.0cm}
\caption[XRT light curves]{
{\bf Left (a--d):} {\it Swift}/XRT 0.2--10\,keV light curves of our sample 
for a {\bf long-term monitoring program} (Table~\ref{fermi11:tab:campaign}, 
2007 October 26 to 2009 November 3). 
Each point refers to the average flux observed
during each observation performed with XRT, except for outbursts 
where the data were binned to 
include at least 20 counts bin$^{-1}$ to best represent the 
dynamical range. Downward-pointing arrows are 3-$\sigma$ upper limits, 
upward pointing arrows mark either outbursts that  
XRT could not observe because the source was Sun-constrained,
or BAT Transient Monitor bright flares. 
AX~J1841.0$-$0536 was only observed during the first year. 
Data from \cite{Romano2009:sfxts_paperV,Romano2011:sfxts_paperVI}.  \\

{\bf left: (e)} 
{\it Swift}/XRT 0.2--10\,keV light curve of IGR~J18483$-$0311 during 
our monitoring program along {\bf one orbital period} (2009 June 11 to 2009 July 9)
at a binning of at least 20 counts bin$^{-1}$. 
 Data from \cite{Romano2010:sfxts_18483}.  \\
{\bf Right:} 
{\it Swift}/XRT light curves of the most representative {\bf outbursts} of SFXTs 
followed by {\it Swift}/XRT,  
referred to their respective BAT triggers
(IGR~J11215$-$5952 did not trigger the BAT, so it is referred to MJD 54139.94).
Points denote detections (binning of at least 20 counts bin$^{-1}$), triangles 3$\sigma$ upper limits.  
%Where no data are plotted, no {\it Swift} data were collected.  
Vertical dashed lines mark time intervals equal to 1 day, up to a week. 
References: 
(f) IGR~J08408--4503   (2008-07-05, \cite{Romano2009:sfxts_paper08408});
(g) IGR~J11215$-$5952  (2007-02-09, \cite{Romano2007}); 
(h) IGR~J16479$-$4514  (2005-08-30, \cite{Sidoli2008:sfxts_paperI}); 
(i) XTE~J1739$-$302    (2008-08-13, \cite{Sidoli2009:sfxts_paperIV}); 
(j) IGR~J17544$-$2619  (2010-03-04, \cite{Romano2011:sfxts_paperVII}); 
(k) SAX~J1818.6$-$1703 (2009-05-06, \cite{Sidoli2009:sfxts_sax1818}); 
(l) AX~J1841.0$-$0536  (2010-06-05, \cite{Romano2011:sfxts_paperVII}). 
Adapted from \cite{Romano2011:sfxts_paperVII}.  \\
}
\label{fermi11:fig:alllcvs}
\end{figure*}%%%%%%%%%%%%%%%%%%%%%%%%%%%%%%%%%%%%%%%%
\begin{figure*}[]%%%%%%%%%%%%%%%%%%%%%%%%%%%%%%%%%%%%%%%
 		\vspace{-4.5truecm}
 		\hspace{-0.5truecm}
\centering
\includegraphics[width=16cm,height=12cm,angle=0]{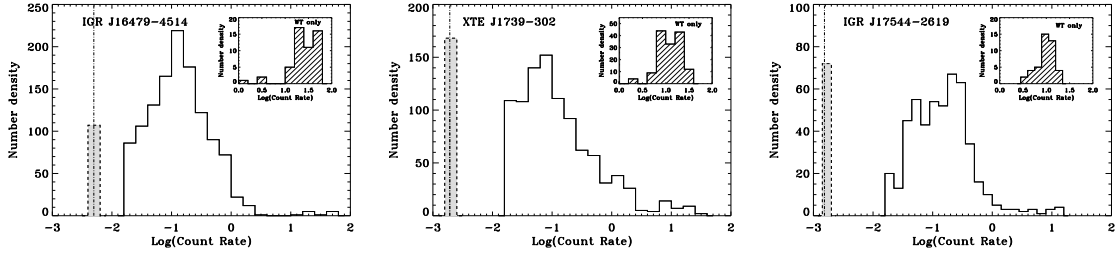}
\vspace{-4.0cm}
                \caption{Distribution of the count rates when the XRT light curves are binned at 100\,s,
for the three sources monitored for two years.  
The vertical lines correspond to the background. The hashed histograms are points which are consistent with 
a zero count rate. The insets show the subset of WT data only, binned at 20\,s.
		}\label{fermi11:fig:histos}
\end{figure*}%%%%%%%%%%%%%%%%%%%%%%%%%%%%%%%%%%%%%%%%%%%

\begin{table*}[t]%%%%%%%%%%%%%%%%%%%%%%%%%%%%%%%%%%%%%%%%%%%%% Table 2 
\begin{center}
\begin{tabular}{lllccccrrc}
\hline
\hline
Name             &Spectrum         & Date/Rate           & $N_{\rm H}$ & $\Gamma$ & $E_{\rm cut}$ & $E_{\rm fold}$  & Flux$^a$ & Luminosity$^b$ & Reference\\           
                 &              & (count s$^{-1}$) &  ($10^{22}$~cm$^{-2}$) &          & (keV)       & (keV)       & (2--10 keV) & (2--10 keV)  &     \\  
\hline
IGR~J16479$-$4514 &outburst &2008-03-19               & $6.2_{-0.5}^{+0.6}$ &$1.2_{-0.1}^{+0.2}$ &$6_{-1}^{+1}$ &$15_{-2}^{+3}$ & 600& 240 & \cite[][This work]{Romano2008:sfxts_paperII} \\
                   & high      & $>$0.55      &$8.2_{-0.7}^{+0.8}$ &$1.1_{-0.2}^{+0.2}$ &    & & 12& 5 & \cite{Romano2011:sfxts_paperVI}        \\ 
                   & medium   & [0.22--0.55[ &$8.6_{-0.8}^{+0.8}$ &$1.3_{-0.2}^{+0.2}$ &    & & 5.3&2 &  \cite{Romano2011:sfxts_paperVI} \\
                   & low      & [0.06--0.22[ &$7.1_{-0.6}^{+0.6}$ &$1.4_{-0.1}^{+0.2}$ &   & & 1.7&0.7 &  \cite{Romano2011:sfxts_paperVI}  \\ 
                   & very low$^{c}$ & $<$0.06 &$3.3_{-0.0}^{+0.4}$  &$1.8_{-0.2}^{+0.3}$ &  & & 0.13&0.04 & \cite{Romano2011:sfxts_paperVI}    \\ 
XTE~J1739$-$302   &outburst &2008-08-13        &$4.8_{-0.6}^{+1.3}$ &$0.8_{-0.2}^{+0.4}$   &$5_{-1}^{+2}$       &$12_{-3}^{+8}$     & 170&18  & \cite[][This work]{Romano2011:TEXAS2010} \\
                    & high     &$>$0.405      &$3.7_{-0.4}^{+0.5}$ &$0.8_{-0.1}^{+0.2}$  &   & & 12&1 & \cite{Romano2011:sfxts_paperVI}    \\ 
                     & medium   &[0.07--0.405[ &$3.8_{-0.4}^{+0.4}$ &$1.4_{-0.1}^{+0.1}$  &  & & 1.8&0.2 & \cite{Romano2011:sfxts_paperVI}     \\ 
                     & very low$^{c}$ &$<$0.07 &$1.7_{-0.0}^{+0.1}$  &$1.4_{-0.2}^{+0.2}$  &  & & 0.05&0.004 & \cite{Romano2011:sfxts_paperVI}     \\ 
IGR~J17544$-$2619 &outburst &2009-06-06        &$1.0_{-0.3}^{+0.2}$  &$0.6_{-0.4}^{+0.2}$   &$3_{-1}^{+1}$ &$8_{-3}^{+4}$ & 83&14  & \cite[][This work]{Romano2011:sfxts_paperVI}\\
                  & high     &$>$0.25       &$1.9_{-0.2}^{+0.3}$ &$1.3_{-0.1}^{+0.1}$ &     & &4.6&0.8 & \cite{Romano2011:sfxts_paperVI} \\ 
                     & medium   &[0.07--0.25[  &$2.3_{-0.3}^{+0.3}$ &$1.7_{-0.2}^{+0.2}$ &  & & 1.4&0.3 &  \cite{Romano2011:sfxts_paperVI}   \\ 
                     & very low$^{c}$ &$<$0.07 &$1.1_{-0.0}^{+0.1}$  &$2.1_{-0.2}^{+0.2}$ &  & & 0.02&0.003 & \cite{Romano2011:sfxts_paperVI}    \\ 
AX~J1841.0$-$0536 &outburst &2010-06-05  &$1.9_{-1.0}^{+1.7}$ &$0.2_{-0.5}^{+0.4}$ &$4_{-4}^{+12}$ &$16_{-9}^{+10}$ &60 &18  & \cite{Romano2011:sfxts_paperVII}  \\
                    & high     & $>$0.4       &$2.5_{-0.3}^{+0.3}$ &$1.1_{-0.1}^{+0.1}$ &   & &8 &3 & \cite{Romano2009:sfxts_paperV}    \\ 
                     & medium   & [0.18--0.4[  &$3.5_{-0.5}^{+0.5}$ &$1.3_{-0.2}^{+0.2}$ &  & &3.4 &1 & \cite{Romano2009:sfxts_paperV}     \\ 
                     & low      & [0.05--0.18[ &$3.5_{-0.5}^{+0.5}$ &$1.5_{-0.2}^{+0.2}$ &  & &1.1 &0.4 & \cite{Romano2009:sfxts_paperV}     \\ 
                     & very low$^{\mathrm{c}}$ & $<$0.05  &$0.3_{-0.3}^{+0.3}$ &$0.6_{-0.4}^{+0.4}$ & & &0.06 &0.02 & \cite{Romano2009:sfxts_paperV} \\ 
\hline
\end{tabular}
\caption{Spectral parameters of {\it (i)}
the outbursts shown in Figure~\ref{fermi11:fig:outburst_spectra} (fit with an absorbed power-law model with a high energy cut-off) 
and {\it (ii)} of the out of outburst states (fit with a simple absorbed power-law model).   
}
  \begin{list}{}{}
  \item[$^{\mathrm{a}}$]{Average observed 2--10\,keV fluxes in units of 10$^{-11}$~erg~cm$^{-2}$~s$^{-1}$.}
  \item[$^{\mathrm{b}}$]{Average 2--10\,keV luminosities in units of 10$^{35}$~erg~s$^{-1}$.  }
  \item[$^{\mathrm{c}}$]{Fit performed with constrained column density.}
  \end{list}  
\label{fermi11:tab:spectra}
\end{center}
\end{table*}%%%%%%%%%%%%%%%%%%%%%%%%%%%%%%%%%%%%%%%%%%%%%

\subsection{SPECTROSCOPIC PROPERTIES \label{fermi11:spectroscopy}}

{\bf Outbursts.} 
Since 2008,  simultaneous observations with XRT and BAT allowed us to perform  
broad band spectroscopy of SFXT outbursts 
\cite{Romano2008:sfxts_paperII,Sidoli2009:sfxts_paperIII,Sidoli2009:sfxts_paperIV,Romano2011:sfxts_paperVI,Romano2011:sfxts_paperVII}. 
Given the shape of the SFXT spectrum (power law with an exponential cut-off at 15--30\,keV), 
the large (0.3--150\,keV) {\it Swift} energy range allows us to both constrain the 
hard-X spectral properties to compare with popular accreting neutron star models 
and to obtain a measure of the absorption. 
Figure~\ref{fermi11:fig:outburst_spectra} shows the spectrum of one outburst of each 
of the four SFXTs 
monitored, while Table~\ref{fermi11:tab:spectra} reports homogenized  
values of the spectral parameters (`outburst' spectra).

{\bf Out of outburst.} 
To characterize the out-of-outburst spectral properties, in each observation we accumulated 
events when the source was not in outburst and a detection was achieved 
 \citep{Romano2009:sfxts_paperV,Romano2011:sfxts_paperVI}.  
We considered several intensity levels (`high', `medium', `low'). 
We also extracted spectra from the event lists accumulated from all observations  
for which no detections were obtained as single exposures  (`very low').
We performed fits in the 0.3--10\,keV energy band with simple models such as an 
absorbed power law or a blackbody as more complex models were not required by the data
(Table~\ref{fermi11:tab:spectra}, `high', `medium', `low', and `very low' spectra; see 
Figure~\ref{fermi11:fig:outburst_spectra}).

\subsection{TIMING PROPERTIES \label{fermi11:timing}}

{\bf Outbursts.} 
Our systematic investigation 
with {\it Swift} (see Figure~\ref{fermi11:fig:alllcvs}f--l 
for the best examples of XRT outburst lightcurves) 
has shown that the common X--ray characteristics of this class include 
outburst lengths well in excess of hours, with a
multiple-peaked structure,  
and a dynamic range (including bright outbursts)  
up to $\sim3$ orders of magnitude.

{\bf Out of outburst.} 
Our monitoring campaigns with {\it Swift} (Figure~\ref{fermi11:fig:alllcvs}a--d) 
have investigated all phases of the SFXT life 
by assessing long each source spends in each state using a systematic 
monitoring with a sensitive instrument. 
The overall dynamic range reaches then $\sim4$ orders of magnitude 
\citep{Romano2009:sfxts_paperV,Romano2011:sfxts_paperVI}. 
We discovered that X-ray emission from SFXTs is still present outside the 
bright outbursts \citep[][see Figure~\ref{fermi11:fig:alllcvs}a--d]{Sidoli2008:sfxts_paperI}, 
that only account for 3--5\,\% of the total lifetime \citep{Romano2011:sfxts_paperVI}.  
Figure~\ref{fermi11:fig:histos} reports the histograms of the XRT count rates, 
peaking at about 0.1 counts s$^{-1}$, which implies that the
most probable X-ray flux is $F_{\rm 2-10\,keV}\sim (1$--$2)\times10^{-11}$ erg cm$^{-2}$ s$^{-1}$ 
(unabsorbed), corresponding to luminosities in the order of a few 10$^{33}$--10$^{34}$~erg~s$^{-1}$, 
$\sim 100$ times lower than the bright outbursts \citep{Romano2011:sfxts_paperVI}. 
Finally, we calculated that the duty-cycle of {\bf inactivity} is in the 
range $\sim 19$--$55$\,\% \citep[Table~\ref{fermi11:tab:campaign};][]{Romano2009:sfxts_paperV,Romano2011:sfxts_paperVI}, 
so that true quiescence is a relatively rare state, 
at variance with what previously estimated using less sensitive instruments. 
{\it Swift} data thus demonstrated that these transients accrete matter through all their life.

{\bf Further monitoring programs} involved following SFXTs with known orbital period. 
In 2009 we performed the first complete monitoring of the X-ray activity along an entire 
orbital period ($P_{\rm orb}\sim 18.5$\,d) of the SFXT IGR~J18483$-$0311 \citep{Romano2010:sfxts_18483} 
with a sensitive instrument (23 daily observations, $\sim2$\,ks each, spread over 28\,d for a total of 44\,ks,
see Fig~\ref{fermi11:fig:alllcvs}e).  
This unique dataset allowed us to constrain the different mechanisms 
proposed to explain the SFXT nature. In particular, 
we applied the clumpy wind model for blue supergiants 
\citep{Ducci2009} to the observed X-ray light curve. 
By assuming an eccentricity of $e = 0.4$, 
we could explain the X-ray emission in terms of the accretion from a spherically 
symmetric clumpy wind, composed of clumps with different masses, 
ranging from $10^{18}$ to $\times10^{21}$\,g.

\section{CONCLUSIONS \label{fermi11:conclusions}}

Thanks to {\it Swift}, we have investigated the properties of SFTXs on 
timescales ranging from minutes to years and in several intensity states 
(bright flares, intermediate intensity states, and down to almost quiescence). 
We also performed broad-band spectroscopy of outbursts, and intensity selected 
spectroscopy outside outbursts. 
In light of their possible emission in the {\it Fermi} energy bands 
SFXTs are certainly worthy of attention at the high energies, especially 
now that {\it Fermi} has a accumulated a long baseline of data and 
has an improved model of the Galactic diffuse emission. 
The spectroscopic and, most importantly, timing properties of SFXTs we 
have uncovered with {\it Swift} could therefore serve as a guide in search for the 
high energy emission from these enigmatic objects.

\begin{acknowledgments}
We acknowledge financial contribution from the agreement ASI-INAF I/009/10/0. 
\end{acknowledgments}



%\bibitem[{\citenamefont{{Sguera} et~al.}(2006)\citenamefont{{Sguera},
%  {Bazzano}, {Bird}, {Dean}, {Ubertini}, {Barlow}, {Bassani}, {Clark}, {Hill},
%  {Malizia} et~al.}}]{Sguera2006}

\bibliography{biblio.bib}

\end{document}